\documentclass[journal]{IEEEtran}
\usepackage{cite}
\usepackage{amsmath,amssymb,amsfonts}
\usepackage{algorithmic}
\usepackage{graphicx}
\usepackage{booktabs}
\usepackage{multirow}
\usepackage{hyperref}
\usepackage{booktabs}
\usepackage{multirow}
\usepackage{graphicx}
\usepackage{romannum}
\usepackage{textcomp}
\usepackage{comment}
\usepackage{tabularx}
\usepackage{setspace}
\usepackage{color}
\usepackage{xcolor}
\usepackage{subcaption}

\ifCLASSINFOpdf
\else
\fi
\graphicspath{{IEEEtran-2/}} 

\begin{document}
%
%
\title{\textcolor{black}{A Single-Channel Consumer-Grade EEG Device for Brain-Computer Interface: Enhancing Detection of SSVEP and Its Amplitude Modulation}}
%
%
%

\author{Phairot Autthasan, Xiangqian Du, Jetsada Arnin, Sirakorn Lamyai, Maneesha Perera, Sirawaj Itthipuripat, \\ Tohru Yagi, Poramate Manoonpong and Theerawit Wilaiprasitporn, ~\IEEEmembership{Member, IEEE}
\thanks {This work was supported by Robotics AI and intelligent Solution Project, PTT Public Company Limited, Thailand Science Research and Innovation (SRI62W1501), and Thailand Research Fund and Office of the Higher Education Commission (MRG6180028).}

\thanks{P. Autthasan, and T. Wilaiprasitporn are with Bio-inspired Robotics and Neural Engineering (BRAIN) Lab, School of Information Science and Technology (IST), Vidyasirimedhi Institute of Science \& Technology (VISTEC), Rayong, Thailand. {\small Corresponding author: theerawit.w@vistec.ac.th}}

\thanks{X. Du and T. Yagi are with Yagi Lab, Department of Mechanical Engineering, Tokyo Institute of Technology, Tokyo, Japan}

\thanks{J. Arnin is with Department of Biomedical Engineering, Mahidol University, Nakhon Pathom, Thailand.}

\thanks{S. Lamyai is with Data Analysis and Knowledge Discovery Laboratory, Department of Computer Engineering, Kasetsart University, Bangkok, Thailand.}

\thanks{M. Perera is with Department of Computer Engineering, Sirindhorn International Institute of Technology, Pathum Thani, Thailand.}

\thanks{S. Itthipuripat is with Department of Psychology, Vanderbilt University, Tennessee and Neurosciences Graduate Program, University of California, San Diego, California USA and Learning Institute, King Mongkut’s University of Technology Thonburi, Bangkok, Thailand.}

\thanks{P. Manoonpong is with BRAIN Lab, IST, VISTEC, Rayong, Thailand and Embodied AI \& Neurorobotics Lab, Centre for BioRobotics, The M{\ae}rsk Mc-Kinney M{\o}ller Institute, The University of Southern Denmark, Odense M, DK-5230, Denmark.}
}

%
%

\markboth{Journal of \LaTeX\ Class Files,~Vol.~14, No.~22, July~2019}%
{Shell \MakeLowercase{\textit{et al.}}: Bare Demo of IEEEtran.cls for IEEE Journals}
%



\maketitle

\begin{abstract}
\textcolor{black}{Brain-Computer interfaces (BCIs) play a significant role in easing neuromuscular patients on controlling computers and prosthetics. Due to their high signal-to-noise ratio, steady-state visually evoked potentials (SSVEPs) has been widely used to build BCIs. However, currently developed algorithms do not predict the modulation of SSVEP amplitude, which is known to change as a function of stimulus luminance contrast. In this study, we aim to develop an integrated approach to simultaneously estimate the frequency and contrast-related amplitude modulations of the SSVEP signal. To achieve that, we developed a behavioral task in which human participants focused on a visual flicking target which the luminance contrast can change through time in several ways. SSVEP signals from 16 subjects were then recorded from electrodes placed at the central occipital site using a low-cost, consumer-grade EEG. Our results demonstrate that the filter bank canonical correlation analysis (FBCCA) performed well in SSVEP frequency recognition, while the support vector regression (SVR) outperformed the other supervised machine learning algorithms in predicting the contrast-dependent amplitude modulations of the SSVEPs. These findings indicate the applicability and strong performance of our integrated method at simultaneously predicting both frequency and amplitude of visually evoked signals, and have proven to be useful for advancing SSVEP-based applications.}

\end{abstract}


%
\IEEEpeerreviewmaketitle

\section{Introduction}\label{sec:introduction}

\IEEEPARstart{B}{RAIN}-Computer Interface (BCI) is a system that reads and converts neuronal activity into an artificial signal that controls computers and machines \cite{847823, wolpaw2002brain, 6712069, 1642776}. BCI is considered to be one of a few promising methods that advance the development of neuroprosthetics for patients with neuromuscular disorders, such as amyotrophic lateral sclerosis, spinal cord injury and brainstem strokes \cite{5342720, Birbaumer, zheng2016unobtrusive}. To implement BCI systems, researchers and developers often use electroencephalography (EEG) to measure brain activity on the scalp, which is generated by the synchronized activity of billions of neurons that lay perpendicularly to the cortical surface \cite{LUCK2013391, LOPESDASILVA20131112}. EEG is one of the most popular measurement methods in BCI research because, it is non-invasive, portable, relatively more affordable compared to other neural measurement techniques, and fast with millisecond precision \cite{Waldert1000}. \textcolor{black}{In the meantime, physiological recordings become a mandatory part of modern medical applications as mentioned in several smart healthcare systems \cite{pirbhulal2018heartbeats,sodhro2018joint,pirbhulal2019medical}. One of those renowned systems is called wireless body sensor network (WBSN), a promising framework for telehealth with great mobility when using off-site. A large number of real-time EEG applications tend to utilize this technology because of its reliability, consistency, and scalability \cite{hussein2015scalable,liu2019fpga}.}


There are at least three types of neural measurements that have been the focus of EEG-based BCI research. These include motor imagery (MI), event-related potential (ERP) and steady-state visually evoked potential (SSVEP) \cite{PFURTSCHELLER19991842, 2, DONCHIN1970201, 847819}. Among these EEG measurements, SSVEP has been widely used in BCI systems, which monitors early sensory processes related to visual stimuli. Using SSVEP to build BCIs is beneficial in numerous ways. Firstly, SSVEP is an early sensory signal oscillating at the exact frequency as the frequency of the incoming visual input \cite{Norcia}, and most importantly, its amplitude changes as a function of stimulus intensity (e.g., luminance contrast) and as a function of attention \cite{kim2007attention, Müller,Müller2, Itthipuripat112, Itthipuripat8635, Itthipuripat6162}. Furthermore, SSVEP is not only a powerful neural index of early sensory processing directly related to a given visual stimulus, but also a method that tracks the attentiveness of human observation at a specific stimulus input. Moreover, SSVEP has a relatively high signal-to-noise ratio (SNR) and information transfer rate (ITR), and a low amount of behavioral training, as well as no requirement of prior experiences in using BCIs \cite{4625401, 5378643}. As a consequence of these qualities, SSVEP has been more favorable compared to other neural measurement techniques mentioned above. 

For decades, numerous researches have focused on developing frequency recognition methods for SSVEP-based BCI. Traditional power spectrum density analyses (PSDA) such as fast Fourier transform (FFT) was used as a conventional method to classify SSVEP frequencies \cite{1642777}. However, the PSDA has a drawback due to its high sensitivity to noise. Accordingly, a state-of-the-art, statistic based SSVEP frequency recognition technique called canonical correlation analysis (CCA) was developed in 2006 and has been proven to be a better method than the PSDA at enhancing the SNR of SSVEP signals \cite{4015614, Bin_2009, Y_lin_mobile}. Afterwards, many research groups have developed hybrid BCI algorithms that could simultaneously detect SSVEP and other EEG measurements such as ERP and event-related desynchronization (ERD) \cite{muller2008control, allison2008towards, luo2010user}. Some of these hybrid methods have been displaying an increase of the information transfer rate (ITR). However, they were still using the traditional frequency analysis approaches, hence the capability of these methods at obtaining high-SNR SSVEP signals might not be as high as it could have been \cite{allison2012hybrid, yin2013novel, yin2014speedy}. Having said that, in 2015, a different research group has proposed the filter bank canonical correlation analysis (FBCCA), which decomposes the SSVEP signals into sub-band components prior to performing the standard CCA procedure, to enhance the efficiency of SSVEP frequency recognition \cite{Chen_2015}. Accordingly, not only did the FBCCA outperformed the traditional PSDA and standard CCA, but it also enhanced the ITR up to 250 bits/min \cite{7740878}. In 2018 to 2019, the learning based-algorithms for SSVEP-BCI have been intensely studied as a novel convolutional neural network (CNN) \cite{zhang2019convolutional}, multiple linear regression models \cite{oikonomou2018bayesian}, deep learning \cite{podmore2019relative}, and as a combined method \cite{ge2019training}. This is an emerging alternative method that is inspired by the world of data-driven applications.

Although the frequency recognition algorithms for SSVEP-based BCIs have been advanced throughout the years of researches as described above, still there is a lack of BCI methods that simultaneously detect the SSVEP frequency and predict amplitude modulations of SSVEP signals, known to change as a function of stimulus intensity. Developing a BCI system that can predict different patterns of SSVEP amplitude modulations across time is highly critical, especially for a smooth control of future neuroprosthetics. In the present study, our research aims at developing an integrated approach to simultaneously estimate the frequency and contrast-related SSVEP-amplitude modulations across time. To do so, we developed a novel visual stimulation protocol, where we presented SSVEP-induced visual stimuli of which luminance contrast remained constant, increased gradually, decreased gradually, and increased and then increased across time. While subjects gazed at these stimuli, we measured SSVEP using an open-source and low-cost consumer-grade EEG device (OpenBCI), placed at the central occipital site. The purpose of using a consumer-grade EEG device was to ensure that our SSVEP analytic approach could be readily applied and developed into a BCI system that will be affordable to consumers. The first step of our approach was the implementation of FBCCA as the SSVEP frequency recognition method, because it has been displayed to yield the best frequency classification and ITR compared to other traditional approaches \cite{Chen_2015, 7740878}. Next, we predicted SSVEP amplitude modulations across time by using linear regression (LR) \cite{Montgomery:Peck:Vining:2007} and different supervised machine learning algorithms including support vector regression (SVR) \cite{Smola2004}, $k$-nearest neighbors ($k$-NN) \cite{KNN} and random forest regression (RF) \cite{Liaw2002} as potential predictive models. 
Filter bank correlation analysis for SSVEP frequency recognition was performed satisfactorily in preceding researches while the support vector regression outperformed the linear regression and the other supervised machine learning algorithms in predicting the time course of contrast-dependent amplitude modulations of the SSVEPs. Together, these results demonstrate the applicability and efficiency of our integrated method at simultaneously predicting both frequency and amplitude of SSVEP signals, and have proven to be useful for advancing consumer-grade SSVEP-based BCIs.

Over the past several years, 
many papers were published focusing mainly on the usage of consumer-grade in a wide variety of BCI applications; for instance, drowsiness detection \cite{mehreen2019hybrid}, age and gender prediction \cite{kaushik2018eeg},  and cognitive assessment \cite{sinha2019readability, zheng2016unobtrusive}. According to our findngs, there were few studies that have reported the advantages of using the consumer-grade EEG toward the SSVEP-BCI \cite{shivappa2018home, wang2018wearable}. For this study, it was of interest to investigate the feasibility of using such EEG devices to enhance the detection of SSVEPs and its amplitude modulations. The experimental results convey four main contributions as follows:

\begin{enumerate}
    \item Feasibility in using the proposed stimulus design to develop EEG-BCI which will utilize both SSVEP frequency and SSVEP amplitude modulations information.
    \item Methodology of EEG data preparation for the proposed predictive model to capture both SSVEP frequency and SSVEP amplitude modulations information, as displayed in \autoref{data_preparation}.
    \item The proposed stimulus design has the ability of guiding the subjects towards contrast-dependent amplitude modulations of the SSVEP signal. 
    \item Possibility in using a single channel EEG with consumer-grade OpenBCI for a future EEG-BCI based on our stimulus design.
\end{enumerate}

The remainder of this paper consists of a section on the data acquisition and two experimental studies (section II). Finally, the results, discussion, and conclusion are contained in sections III, IV and V, respectively.

\section{Materials and Methods}
This section explains the experimental protocol for acquiring SSVEP responses (the dataset) with contrast-dependent amplitude modulations of the SSVEP signal along with the information on EEG data recording with feature extraction. The EEG data recording was analyzed in two studies. The first study was used to classify SSVEP frequency information with a frequency recognition method, while the second study was used to construct a predictive model for predicting amplitude modulations of the SSVEP signal.

\subsection{Data acquisition}
The participants of this experiment were sixteen healthy people aged between 20 and 34 ($n = 16$) with normal or corrected-to-normal vision. The experiments received approved consent from all participants following the Helsinki Declaration of 1975 (as revised in 2000), which was approved by the internal review board of Rayong Hospital, Thailand.

\subsubsection{EEG data recording}
The typical SSVEP studies have mainly relied on the usability of visual stimuli with a fixed frequency to generate SSVEP signals at that frequency in the visual cortex area of the brain \cite{10.1167/15.6.4,6346301}. We usually obtain these SSVEP signals from occipital and parietal areas. $O_{z}$ is a position for a high-density recording, which lies in the mid-line sagittal plane, and usually gives the highest SSVEP magnitude \cite{10.3389/fncir.2013.00027, 6098285}. 
\textcolor{black}{In this study, an open-source, low-cost, and consumer-grade EEG device, namely \textit{OpenBCI} \cite{OpenBCI}, was used with sampling frequency of 250 Hz. The device is developed based on the Analog Front-End (AFE) ADS1299 (Texas Instruments, USA) \cite{s18113721}. Recently, many studies published in high reputation journals have focused on the usability of OpenBCI device in an numerous variety of BCI applications \cite{app_openbci1, app_openbci2, app_openbci3}.} For practical purposes, a single-channel EEG ($O_{z}$) was used to record data
throughout all the experiments. Electrode impedance were kept less than 5 k${\Omega}$. The reference and ground electrodes were placed at both of the earlobes.

\subsubsection{Stimulation protocol}

\begin{figure}
\includegraphics[width=1.0\columnwidth]{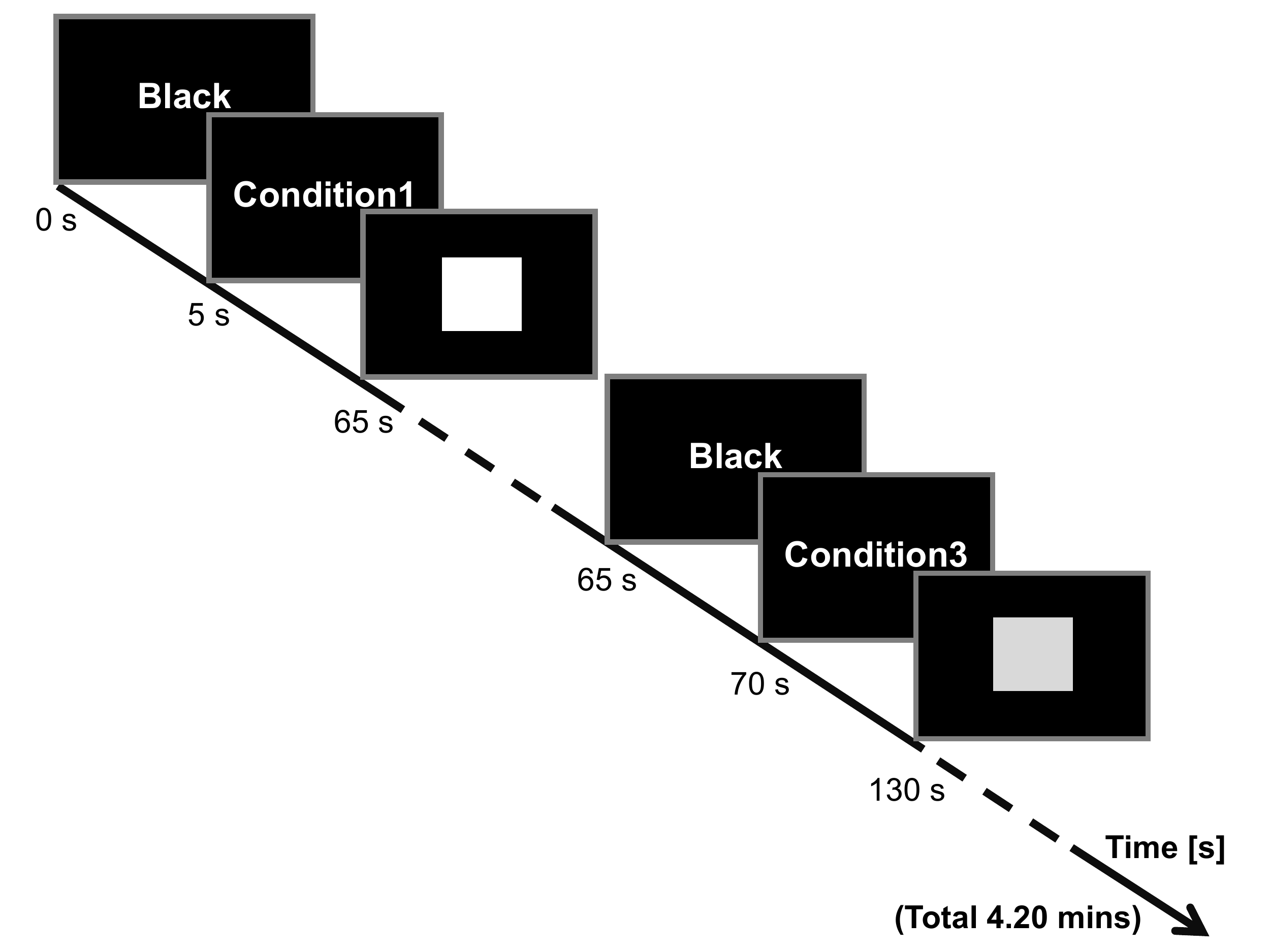}
\caption{Four stimulus conditions presented randomly to the subjects, each lasting for 60 seconds. A black screen and a conditional cue were both displayed for 5 seconds each before the beginning of every condition.}
\label{protocol}
\end{figure}

To ensure the practicality of the study outcomes in continuing development of real-world applications, the experiments were conducted in a normal environment (a room without electromagnetic shielding). The subjects were asked to sit in front of a 17-inch monitor, put their heads on a chin-rest position 30 cm away from the screen, and pay constant attention to the center of the screen. \autoref{protocol} illustrates the SSVEP stimulus protocol. Four stimulus conditions with a frequency of 7.5 Hz were presented in a random order to the subjects. This frequency was selected as the target frequency because, it has the highest SSVEP amplitude in empirical tuning. Afterwards, we provided the conditions and each condition lasted for 60 seconds. A black screen and a conditional cue were both displayed for 5 seconds each, before the beginning of every condition. The conditions were as follows:

\begin{enumerate}
\item A 270 px $\times$ 270 px black/white square flickering at 7.5 Hz is in the center of the screen. It is constantly flickering at maximum luminance contrast. \textit{Note: Luminance contrast can vary from level 0 (minimum contrast) to level 255 (maximum contrast).} This condition is the conventional SSVEP stimulation and serves as the baseline for other conditions (cond.1). 
\item The same square starts flickering at a contrast level of 75. The luminance contrast is then increased gradually by three levels per second for 60 seconds. This condition is supposed to stimulate the subjects to increase their SSVEP magnitude (cond.2).
\item The same square starts flickering at the maximum contrast (255). The luminance contrast is then decreased gradually  by three levels per second for 60 seconds. This condition is supposed to stimulate the subjects to decrease their SSVEP magnitude (cond.3).
\item The square starts flickering at the contrast level of 150. For the initial 30 seconds of 60 seconds time period, the contrast is increased gradually by three levels per second until it reaches maximum. For the next 30 seconds, the luminance contrast is decreased gradually by three levels per second until the end of the condition. This condition is supposed to stimulate the SSVEP magnitude of subjects by increasing and decreasing inside the same condition (cond.4).
\end{enumerate}

The stimulus program was developed using Processing software version 3.4 \cite{processing}. The visual stimulation of a contrast level at \texttt{x} is done by the color filling of the stimulation target with the command \texttt{color(x);} in the Processing. It should be observed that the maximum contrast level of 255 will produce a white color stimulus, which gives the maximum contrast to the black background, while the level of 0 gives a black color stimulus which is identical to the background. 

To prepare the dataset for the remainder of the study, a notch filter at 50 Hz (to filter out electrical noise) and a band-pass filter (Butterworth, order 2) at 6-25 Hz (to cover three harmonics of the target stimulus: 7.5, 15, and 22.5) were applied on the EEG data. \textcolor{black}{All filter techniques were based on typical pre-processing for SSVEP-EEG.} The SSVEP responses were obtained from the filtered signals and then segmented according to the experimental conditions. Eventually, 60 seconds long SSVEP responses from sixteen subjects were obtained for each condition.

\subsection{Investigatory structure towards an integrated approach to simultaneously estimate the frequency and amplitude modulations of SSVEP signals}
\begin{figure*}
\centering
\includegraphics[width=2.00\columnwidth]{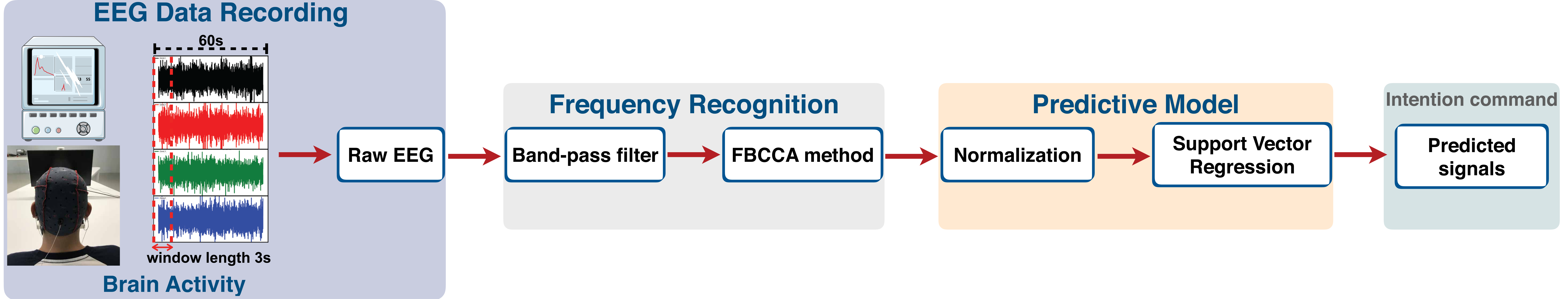}
\caption{\textcolor{black}{The design of the structure for simultaneously estimating the frequency and amplitude modulations of SSVEP signals.}}
\label{Schematic of system}
\end{figure*}

\begin{figure}
\centering
\includegraphics[width=1\columnwidth]{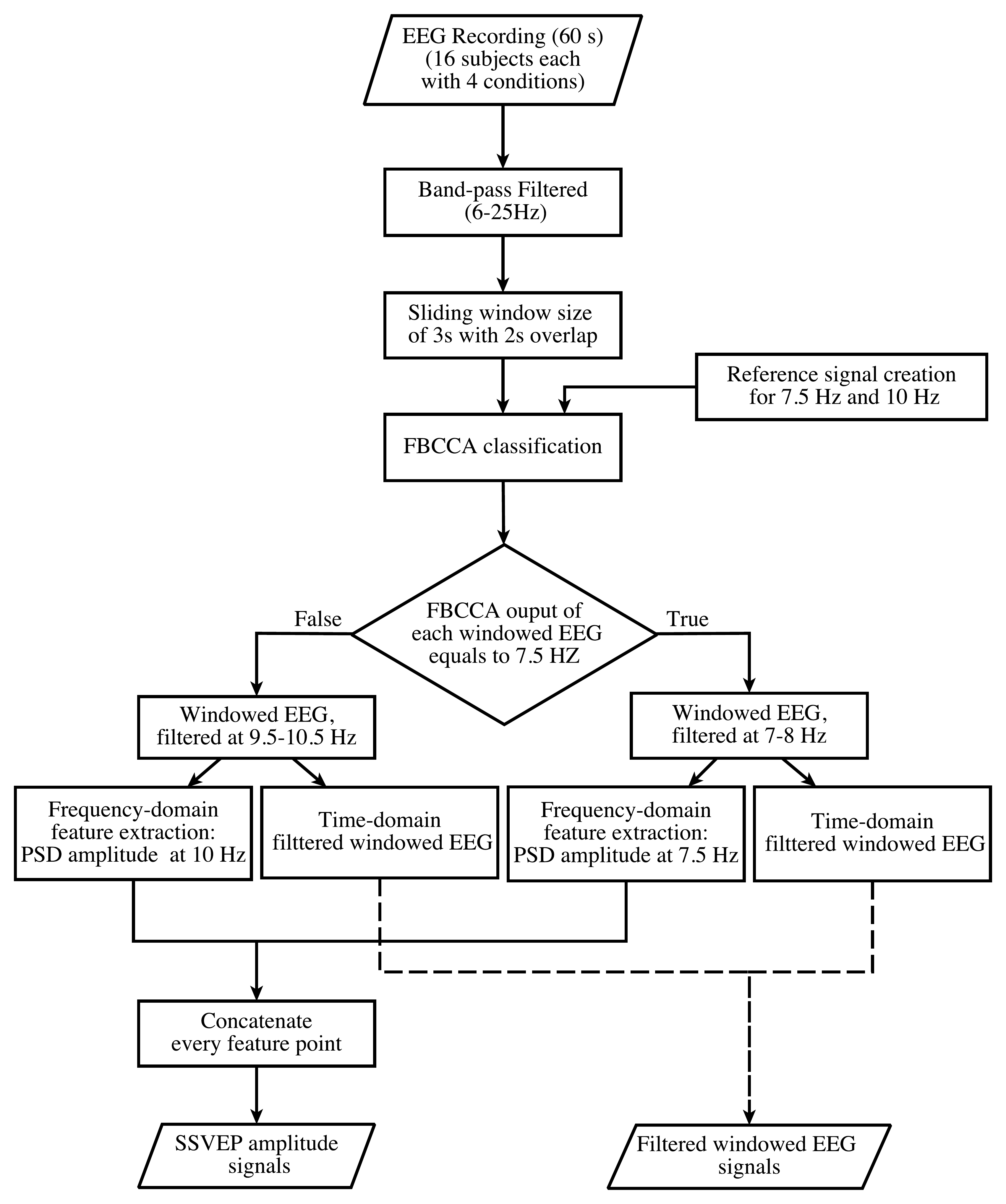}
\caption{\textcolor{black}{Flowchart of processing data for the proposed predictive model to capture both SSVEP frequency and SSVEP amplitude modulations information.}}
\label{data_preparation}
\end{figure}

\begin{figure*}
\centering
\includegraphics[width=2.03\columnwidth]{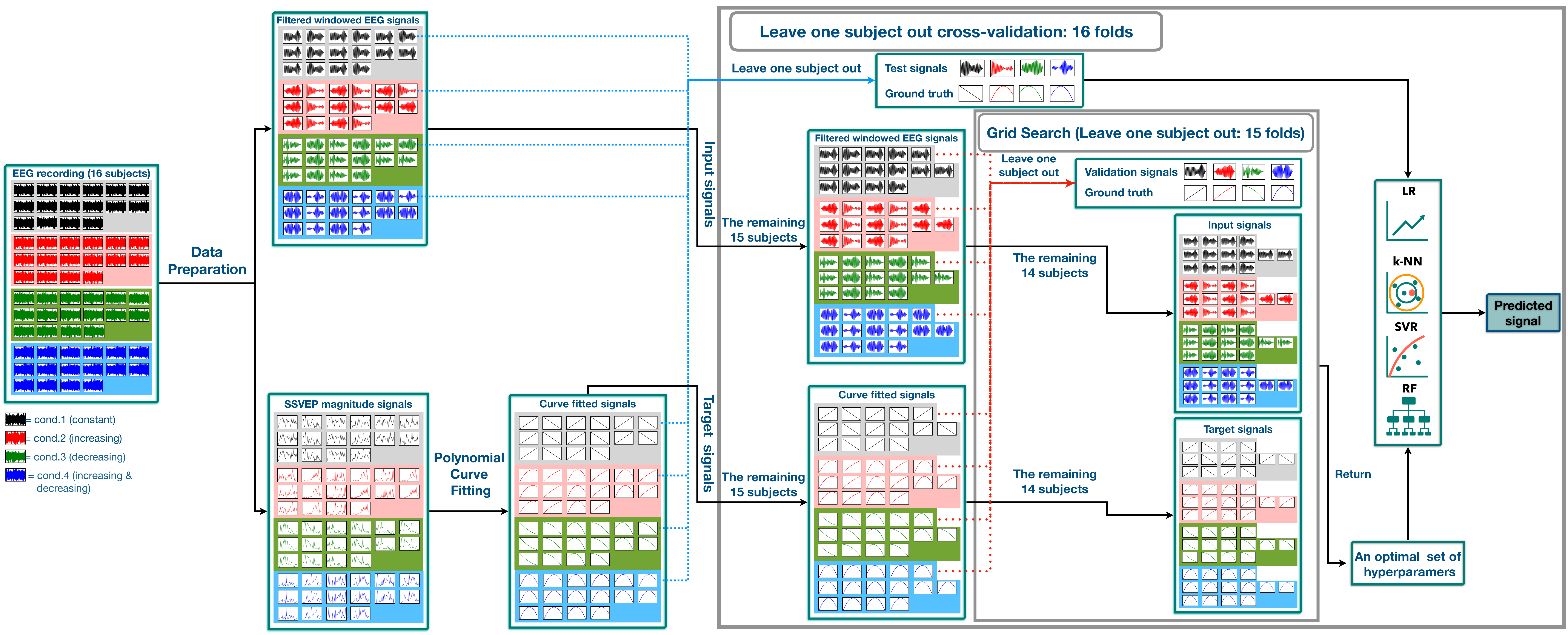}
\caption{\textcolor{black}{Architecture of leave-one-subject-out cross-validation (LOCV) with the grid search algorithm for training the model.}}
\label{predictive_model}
\end{figure*}

\autoref{Schematic of system} displays the structure for an integrated approach to simultaneously estimate the frequency and amplitude modulations of the SSVEP signal based on utilizing both SSVEP frequency and SSVEP amplitude modulations information. This structure mainly consists of SSVEP-EEG data recording, frequency recognition, and predictive model. The designed structure will enable the possibility of using both the SSVEP frequency and SSVEP amplitude modulations information for handling the intended response. Firstly, the EEG recording was used to acquire SSVEP responses from 16 subjects by presenting the SSVEP stimulus protocol as \autoref{protocol}. Subsequently, the frequency recognition was applied for classifying SSVEP frequency information, while the system allows the user to focus on the target stimulus. Finally, the predictive model was operated in predicting amplitude modulations of the SSVEP signal to translate SSVEP amplitude modulations information to intended commands (machine command).     

\subsection{Experiment I: Frequency recognition for SSVEP-based BCI}
\label{exp1}
\textcolor{black}{Main focuses of this experiment were twofold: (1) to examine the frequency recognition of SSVEP signals and (2) to find the optimal processing window length according to the selected frequency recognition method.} Alpha band activity in human EEG is suppressed by stimulating with a visual stimulus and it is approximately dominated in the range of 8 to 12 Hz. In our study we used the center frequency of the above mentioned range (10 Hz) to reflect the resting state \cite{KLIMESCH2012606,van}. We applied the standard CCA and FBCCA methods to distinguish between the target stimulus (7.5 Hz) and resting state (10 Hz) for all subjects in each condition of our dataset.

CCA is the most common statistical method which is used to detect the frequency of SSVEPs \cite{4015614}. This can be used to measure the fundamental correlation between two multi-dimensional variables. The goal is to maximize the correlation between these two multi-dimensional variables. Considering the linear combinations $\boldsymbol{x} = \boldsymbol{X}^T \boldsymbol{W_X}$ and $\boldsymbol{y} = \boldsymbol{Y}^T \boldsymbol{W_Y}$, where $\boldsymbol{X}$ and $\boldsymbol{Y}$ are the multidimensional variables, standard CCA seeks the weight vectors $\boldsymbol{W_X}, \boldsymbol{W_Y}$ and maximize the correlation between $\boldsymbol{x}$ and $\boldsymbol{y}$ using the following formula:
\begin{equation}
\small
\underset{\boldsymbol{W_X},\boldsymbol{W_Y}}{\max}
{\rho(\boldsymbol{x}, \boldsymbol{y})} = 
\frac
{E \left[ \boldsymbol{W}^T_{\boldsymbol{X}} \boldsymbol{X} \boldsymbol{Y}^T \boldsymbol{W_Y} \right]}
{\sqrt{E \left[ \boldsymbol{W_X}^T\boldsymbol{XX}^T\boldsymbol{W_X}\right]\left[ \boldsymbol{W_Y}^T\boldsymbol{YY}^T\boldsymbol{W_Y}\right] }}
\end{equation}
The maximum $\rho$ with respect to $\boldsymbol{W_X}$ and $\boldsymbol{W_Y}$ gives the maximum canonical correlation. When detecting the frequency in SSVEPs, $\boldsymbol{X}$ and $\boldsymbol{Y}$ refer to indicate the multi-channel SSVEPs and sine-cosine reference signals respectively. These reference signals have the same length as $\boldsymbol{X}$. The reference signals $\boldsymbol{Y}_f$ are given by
\begin{equation}
\small
\boldsymbol{Y}_f=
\begin{bmatrix}
\sin(2\pi
fn)\\
\cos(2\pi
fn)\\
\vdots\\
\sin(2\pi
N_hfn)\\
\cos(2\pi
N_hfn)\\
\end{bmatrix}, \quad
n = \frac{1}{f_s},\frac{2}{f_s},\dotsc,\frac{N_s}{f_s}
\end{equation}
where $f$ is the stimulation frequency, $N_h$ is the harmonic number, $f_s$ is the sampling rate, and $N_s$ is the amount of sampling points. To identify the frequency of SSVEPs, CCA calculates the canonical correlation between multi-channel SSVEPs and the sine-cosine reference signals corresponding of each stimulation frequency, and find the maximum correlation between them. The frequency of reference signals with the maximum correlation is the frequency of SSVEPs.

To enhance the standard CCA-based frequency detection of SSVEPs, a research group has proposed the idea of filter bank CCA (FBCCA) \cite{Chen_2015}. This method consists of three main procedures: filter bank analysis, CCA between SSVEP sub-band components and sinusoidal reference signals, and target identification. In the first step, the filter bank analysis is performed. Here, we project the original EEG signals $\boldsymbol{X}$, into an array of band-pass filters which separates the input signal into multiple components. These components carry a single frequency sub-band ($\boldsymbol{X}_{\textrm{SB}_n}, n=1, 2, \dotsc, N$) of the original signal. In the second procedure, the standard CCA process is applied to each of the sub-band component separately, which leads to a correlation between multi-channel SSVEPs and the sine cosine reference signals ${\boldsymbol{Y}_f}_k$. The correlation vector $\rho_k$ is defined as follows:
\begin{equation}
    \boldsymbol{\rho}_k = 
    \mbox{\small$
    \begin{bmatrix} 
    \rho_k^1\\
    \rho_k^2\\
    \vdots\\
    \rho_k^N\\ 
    \end{bmatrix}$} =
    \mbox{\scriptsize$
    \begin{bmatrix}
    \rho \left( \boldsymbol{X}^T_{\textrm{SB}_1} \boldsymbol{W_X} \left( \boldsymbol{X}_{\textrm{SB}_1} \boldsymbol{Y}_{f_k} \right) , {\boldsymbol{Y}^T} \boldsymbol{W_Y} \left( \boldsymbol{X}_{\textrm{SB}_1} \boldsymbol{Y}_{f_k} \right) \right)\\
    \rho \left( \boldsymbol{X}^T_{\textrm{SB}_1} \boldsymbol{W_X} \left( \boldsymbol{X}_{\textrm{SB}_1} \boldsymbol{Y}_{f_k} \right) , {\boldsymbol{Y}^T} \boldsymbol{W_Y} \left( \boldsymbol{X}_{\textrm{SB}_1} \boldsymbol{Y}_{f_k} \right) \right)\\
    \vdots \\
    \rho \left( \boldsymbol{X}^T_{\textrm{SB}_1} \boldsymbol{W_X} \left( \boldsymbol{X}_{\textrm{SB}_1} \boldsymbol{Y}_{f_k} \right) , {\boldsymbol{Y}^T} \boldsymbol{W_Y} \left( \boldsymbol{X}_{\textrm{SB}_1} \boldsymbol{Y}_{f_k} \right) \right)\\
    \end{bmatrix}$}
\end{equation}
where $\rho(x,y)$ indicates the correlation coefficient between $x$ and $y$. $\boldsymbol{W_X}\left(\boldsymbol{X}_{\textrm{SB}_i}\boldsymbol{Y}_{f_k}\right)$ and $\boldsymbol{W_Y}\left(\boldsymbol{X}_{\textrm{SB}_i}\boldsymbol{Y}_{f_k}\right)$, $i=1, 2, \dotsc, N$ are the linear combination coefficients which were obtained using the standard CCA between $\boldsymbol{X}_{\textrm{SB}_i}$ and $\boldsymbol{Y}_{f_k}$. The weighted sum of all corresponding sub-band components ${\overset{\sim}{\rho_k}}$ are calculated and used as the feature for identification of the target:
\begin{equation}
\overset{\sim}{\rho_k}=\sum_{n=1}^{N} \boldsymbol{w}\left(n\right)\cdot\left(\rho_k^n\right)^2
\end{equation}
where $n$ is the index of the sub-band. The weights for the sub-band components are defined as follows:
\begin{equation}
\boldsymbol{w} \left(n\right) = n^{-a} + b,\quad n \in [1,N]
\end{equation}
where $a$ and $b$ are constants that maximize the classification performance. The constants $a$ and $b$ were determined using a grid search method in an offline analysis. All the stimulated frequencies which are corresponding to ${\overset{\sim}{\rho_k}}$ were used to determine the frequency of SSVEPs. The frequency of the reference signal which has maximized ${\overset{\sim}{\rho_k}}$ is considered to be the SSVEPs frequency.

\textcolor{black}{In this experiment, single-channel EEG ($O_{z}$) obtained from sixteen subjects were used as the input for both the standard CCA and FBCCA; each stimulus condition, the signal was processed by a three-second sliding window shifted in steps of a second. For the FBCCA, $N_h$ was varied from 1 to 3 and then selected the highest classification accuracy from them. 
It is to be observed that this study used $N_h=3$ for all standard CCA and FBCCA methods. All the other parameters in the FBCCA method were identical to the original settings of the feature extraction and frequency classification method that were firstly proposed by \cite{Chen_2015}. Based on this approach, it is supposed that the FBCCA has more promising and higher classification accuracy than the standard CCA. To compare the results from these two methods, a paired t-test was used to determine the difference in the classification accuracy. Besides, with lengths of one, three, five and seven seconds were considered to examine the optimal length of the processing window. However, only steps of 0.5 seconds were used for the one-second processing window, whereas the other window lengths used steps of a second. Subsequently, the data from all processing lengths were evaluated by using the most suitable frequency recognition method. We explored these effects statistically by using the one way repeated measures analysis of variance (ANOVA) based on the assumption of sphericity. Likewise, data correction was applied to the analysis when it violated this assumption so the Bonferroni correction and the pairwise comparison were performed for post hoc analysis.}

\subsection{Experiment II: Predictive models for predicting SSVEP amplitude modulations}

\subsubsection{Data Preparation of EEG}
The flowchart of data preparation process for the proposed predictive model, which was used to capture both SSVEP frequency and SSVEP amplitude modulations information, is displayed in \autoref{data_preparation}. By considering the results of Section \ref{freq_recog},
the FBCCA method was found to be the most appropriate frequency recognition approach in terms of high performance for classifying SSVEP frequency. Meanwhile, the processing window length of three-second (two-second overlap) with a step of a second is considered to be the most appropriate length. Thus, we selected only the FBCCA method with a three-second processing window to extract features from the filtered SSVEP signals. In this way, the process in \autoref{data_preparation} consists of two feature extraction pathways, namely SSVEP amplitude signals and filtered windowed EEG signals. Firstly, the filtered SSVEP signals which were acquired from 16 subjects (each with 4 conditions) were converted into a sequence of sub-samples or sliding windows with a three-second processing window (3 s $\times$ 250 Hz = 750 data points per a windowed EEG) and a two-second overlap. The FBCCA was then used to decompose EEG signals and extract SSVEP frequency information for SSVEP frequency classification. The frequency of these windowed EEGs are classified as 7.5 Hz (7.5 Hz is target stimulus) or 10 Hz (10 Hz is resting state) using the FBCCA method. In order to extract frequency-domain information from the windowed EEG, it is required to apply the band-pass filter. If FBCCA classifies the frequency as 7.5 Hz, a band-pass filter of 7-8 Hz is being applied on that windowed EEG. Otherwise, a band-pass filter of 9.5-10.5 Hz was applied. The time-domain filtered windowed EEGs were then transformed into frequency-domain using the power spectral density (PSD) based on Welch's method \cite{Welch}. Highest PSD amplitudes at 7.5 Hz and 10 Hz were picked from the filtered windowed EEGs and were classified as 7.5 Hz and as 10 Hz respectively. Here, PSD with the related processing was used to calculate the ground truth to train the model during the calibration/training process only. There is no need to perform this step in the online evaluation. Afterwards, every feature point from each filtered windowed EEG was concatenated together and were used to acquire SSVEP amplitude signals. Furthermore, the time-domain filtered windowed EEGs were combined together and were used as filtered windowed EEG signals. After the data preparation was completed, the SSVEP amplitude signals were in a dimension of (16 subjects $\times$ 4 conditions $\times$ 58 feature points) or 64 curves with 58 points each, while the dimension of the filtered windowed EEG signals was (16 subjects $\times$ 4 conditions $\times$ 58 windows $\times$ 750 feature points).

\subsubsection{Predictive models structure}
Generally, predictive models are a common kind of machine learning models which are being used widely to predict a target value on a set of input values. In this study, the proposed predictive model was trained using time-domain SSVEP responses from the human's brain as input signals. Meanwhile, curve-fitted signals from applying curve fitting method on SSVEP amplitude modulations information were used as target signals. In this part, we aim at demonstrating the advantages of the proposed predictive model to predict amplitude modulations of the SSVEP signal. Since we aim to capture the SSVEP amplitude modulations information, the proposed predictive model was only used to manage the regression tasks and there is no classification tasks in this experimental scheme. The machine learning models including random forest regression (RF), $k$-nearest neighbor ($k$-NN) and support vector regression (SVR) were compared in this study. To demonstrate the advantages of machine learning, a simple linear regression (LR) was used as a baseline. LR is a well-known simple predictive model to find a linear relationship between a target value for one or more input values. RF is an ensemble learning method, combining the predictions of multiple smaller decision trees and the final prediction is performed by averaging the results from the decision trees, which tends to reduce overfitting. $k$-NN algorithm stores all the available cases and predict the numerical target based on $k$ data points with the least input norm. SVR fits as many as possible instances while limiting margin violation, rather than fitting the largest possible street while limiting margin violations. Out of the aforementioned machine learning algorithms, SVR method had the most significant results compared to others. Therefore, it is better to explore further details on SVR.

SVR method was derived from SVM which can be used in solving regression problems\cite{Smola2004}. SVR can be used in both linear and non-linear problems. Due to its attractive features and promising empirical performance, it has been gaining popularity over the time. SVR algorithm focuses on generalized error bound minimization and by doing that it controls the overfitting problem. This can be done by,
\begin{equation} \label{eq:erl}
\frac{1}{2}\mathbf{w}^T\mathbf{w} + C \sum_{i=1}^{n}\xi_i + C \sum_{i=1}^{n}\xi^*_i,
\end{equation}
under the constraints of
\begin{equation}
\label{eq:2}
 \begin{array}{l}
\mathbf{w}^T K(x_i)+b-y_i \leq \epsilon + \xi_i,  \\
y_i - \mathbf{w}^T K(x_i)-b \leq \epsilon + \xi^*_i, \\
\textrm{and } 
\xi_i,\xi^*_i\geq0,i=1, \dotsc, n.
\end{array}
\end{equation}
when $C$ is the capacity constant, $\mathbf{w}$ is the vector of coefficients, $b$ is a bias offset, $\epsilon$ is the margin of tolerance, and $y_i$ represents the label of the $i$\textsuperscript{th} training example from the set of $N$ training examples. $\xi_i, \xi^*$ are positive slack variables. The larger the $C$ value, the more the error is penalized. The $C$ value is optimized to avoid overfitting using the validation set, which is described in Section \ref{model_validation}. Lagrange multipliers with Karush-Kuhn-Tucker conditions are applied in a dual problem \eqref{eq:2} to solve for $\mathbf{w}$. The prove to a final solution can be found at \cite{Lin2007}. 
$K(x_i)$, or kernel function is a space-transform function that maps $x_i$ to a higher space dimension. Two kernels were implemented in this comparison:
\begin{itemize}
\item Polynomial kernel (Poly): $K(x_i,x_j) = (1+x_i x_j)^d$
\item Gaussian kernel (RBF): $K(x_i,x_j) = \exp(\frac{-\left\lVert x_i-x_j \right\rVert ^2}{2\sigma^2})$
\end{itemize}

\subsubsection{Model Validation}
\label{model_validation}

\begin{table}
    \caption{List of hyperparameters tuned for all predictive models using grid search}
    \label{hyperparameters}
    \centering
    \resizebox{0.9\columnwidth}{!}{
        \begin{tabular}{@{}cccc@{}}
            \toprule[0.2em]
            \textbf{Model} & \textbf{Parameter} & \textbf{Kernel} & \textbf{Values} \\ \midrule[0.1em]
            SVR & $C$ & all & 0.001, 0.01, 0.1, 1, 10, 100 \\ 
             & $\gamma$ & RBF & 0.01, 0.1, 1, 10, 100 \\ 
             & $\epsilon$  & all & 0.001, 0.01, 0.1, 1, 10 \\ 
             & $d$ & Poly & 0, 1, 2, 3, 4, 5, 6 \\ \midrule[0.07em]
            $k$-NN & Number of neighbors & - & 1-50 with a step of 1 \\ 
             & Type of weights & - & \textit{uniform},  \textit{distance} \\ \midrule[0.07em]
            RF & Max depth & - & 1-32 with a step of 2 \\ 
             & Number of estimators & - & 1, 5, 10, 15, 20 \\
             & Max features &-& 1, 3, 5, 7, 9 \\
             & Min samples split & - & 0.1, 0.3, 0.5, 0.7, 0.9, 1.0  \\
             & Min samples leaf & - & 0.1, 0.3, 0.5 \\ \midrule[0.07em]
            LR & - & - & - \\ 
            \bottomrule[0.2em]
        \end{tabular}
    }
\end{table}

LR, RF, $k$-NN and SVR were considered as predictive models to predict the SSVEP amplitude modulations. Each predictive model and kernel have different hyperparameters which are tuned to get optimal parameters to be used in the model. In order to train the predictive models, as displayed in \autoref{predictive_model}, the EEG data was first prepared into the SSVEP amplitude and the filtered windowed EEG signals as following in \autoref{data_preparation}. Afterwards, input signals and target signals are generated which are described as follows: 

\textbf{Input signals:} In this study, the input signals were constructed using the filtered windowed EEG signals as demonstrated in \autoref{predictive_model}. Due to the concerned relevant feasibility of a future online/real-time control applications, the EEG data had been evaluated by considering to predict a target point in every second. Therefore, the input signals were in the dimension of (16 subjects $\times$ 4 conditions $\times$ 58 windows $\times$ 750 feature points)

\textbf{Target signals:} Here, the target signals were built using the SSVEP amplitude signals as displayed in \autoref{predictive_model}. Curve fittings were performed for each condition with polynomial functions (a linear function (\textit{poly1}) for condition 1 and a quadratic function (\textit{poly2}) for the rest of the conditions). Eventually, the results from curve fitting method (curve-fitted signals) were used as target signals with a dimension of (16 subjects $\times$ 4 conditions $\times$ 58 feature points). Towards practicality in the same way as the input signals, the target signals also considered each feature point in each condition as one sample. 

All predictive models were implemented with leave-one-subject-out cross-validation (LOCV) on sixteen subjects (16 folds) using Scikit-learn \cite{scikit-learn}, as displayed in \autoref{predictive_model}. Each fold consisted of fifteen subjects as the training set and the remaining subject as the testing set. Due to the LOCV had been used in this study, so far this design of structure supported for new users in no requiring calibration method. The input and target signals of training set in each condition have been normalized individually with min-max normalization to scale them into a common range. Besides, the min-max normalized factor from the training set had also been applied to the testing set. In training session, we implemented a hyperparameters optimization algorithm, namely grid search \cite{Bergstra12randomsearch}, to perform the tuning of hyperparameters of all models. The training set was used to establish the optimal set of hyperparameters, which provides the predictive models to return the best mean absolute error (MAE). By considering the grid search algorithm, there was also leave-one-subject-out cross-validation (15 folds). The list of tuned hyperparameters for each model is indicated in \autoref{hyperparameters}. Finally, a predictive model, for each fold, among four proposed predictive model with optimal hyperparameters was evaluated by performing prediction on testing set. To compare these four approaches, the one way repeated measures analysis of variance (ANOVA) was used for statistical analysis. Correction was applied when the data violated the sphericity assumption. Bonferroni correction and pairwise comparison were performed for post hoc analysis.


\section{Results}

\textcolor{black}{This section summarizes the findings and contributions made from each experiment separately. Result I reports a comparative study of the standard CCA and FBCCA methods for SSVEP frequency recognition; the optimal length of the processing window is also reported to manifest the feasibility of classifying SSVEP frequency.} Finally, Result II leads us to the utilization of SSVEP amplitude modulations, which demonstrates the comparative performance of predictive models for predicting amplitude modulations of the SSVEP signal.

\subsection{Result I: Frequency recognition for SSVEP-based BCI}
\label{freq_recog}
\textcolor{black}{This experiment aims at finding the best frequency recognition method for classifying SSVEP frequency. The significance testing was done based on the paired sample t-test so a comparision between the standard CCA approach and the FBCCA approach is reported in this section. \autoref{Comparison_CCAmethod} displays the classification accuracy of both the standard CCA and the FBCCA methods when performed using the three-second sliding window with an overlap of two seconds. With regards to the results, it displays that the FBCCA method outperforms the standard CCA method for all experimental conditions. The paired t-test revealed statistically significant differences in the classification accuracy of two different approaches throughout all experimental conditions, $p < 0.05$. As a result, we finally selected the FBCCA method for performing the frequency recognition of SSVEP-based BCI, which is illustrated in our proposed processing pipeline.}

\textcolor{black}{Previous studies have emphasized the importance of selecting a length of processing window, especially when real-time applications are required. In our system, the incoming EEG data were processed using a sliding window with steps of a second to classify the target frequency every second. Regarding the study of finding the optimal length of the processing window, the effectiveness of the window length was considered from the performance of the FBCCA method. \autoref{Comparison_time_windows} illustrates the average classification accuracy obtained from each experimental condition with different processing window lengths. The univariate repeated measures ANOVA with Greenhouse-Geisser correction displayed a statistical difference in the accuracy among the processing window lengths of the experimental condition 3, $F(1.817, 27.253) = 2.982$, $p=0.072$, while the other conditions found no significant differences. In particular, the pairwise comparison of the condition 3 revealed that the accuracy of the five-second sliding window is significantly higher than the one-second window, $p<0.05$, but did not display significant differences for three and seven seconds. Interestingly, considering the results from the experimental condition 3, with using the three-second processing window length, the results confirm that this is a good choice for being the optimal length for further studies. Moreover, the three-second processing window can maintain the accuracy of the FBCCA without significant differences compared to the other large window lengths.}

\begin{table}[]
\centering
    \caption{\textcolor{black}{Average of binary classification accuracy (n = 16) between the target stimulus (7.5 Hz) and the resting state (10 Hz) of the standard CCA and FBCCA methods across different experimental conditions when processing window length of 3 seconds with a step of 1-second was selected (bold is higher). * Denotes that the number is significantly higher than the others, $p < 0.05$}
    }
    \label{Comparison_CCAmethod}
\footnotesize
\resizebox{0.6\columnwidth}{!}{%
\begin{tabular}{@{}ccc@{}}
\toprule[0.2em]
\multirow{2}{*}{\textbf{\begin{tabular}[c]{@{}c@{}}Experimental\\ Conditions\end{tabular}}} & \multicolumn{2}{c}{\textbf{Accuracy $\pm$ SE}} \\ \cmidrule[0.1em](l){2-3} 
 & \textbf{CCA} & \textbf{FBCCA} \\ \midrule[0.1em]
Cond.1 & 93.32 $\pm$ 1.63 & \textbf{99.25 $\pm$ 0.22*} \\ 
Cond.2 & 88.25 $\pm$ 5.39 & \textbf{99.78 $\pm$ 0.15*} \\ 
Cond.3 & 91.16 $\pm$ 3.21 & \textbf{99.89 $\pm$ 0.11*} \\ 
Cond.4 & 91.16 $\pm$ 2.58 & \textbf{99.78 $\pm$ 0.15*} \\ \bottomrule[0.2em]
\end{tabular}%
}
\end{table}

\begin{table}[]
\centering
    \caption{Average of classification accuracy (n = 16) in each experimental condition for the FBCCA method corresponding to different processing window lengths from 1-second to 7-second (bold is higher). A step of 0.5-second was used for the length of 1-second and a step of 1-second was used for the other lengths. * Denotes that the number is significantly higher than the others, $p<0.05$. 
    }
\label{Comparison_time_windows}
\footnotesize
\resizebox{1\columnwidth}{!}{%
\begin{tabular}{@{}ccccc@{}}
\toprule[0.2em]
\multirow{2}{*}{\textbf{\begin{tabular}[c]{@{}c@{}}Window \\ length {[}s{]}\end{tabular}}} & \multicolumn{4}{c}{\textbf{Avg. Accuracy {[}\%{]} $\pm$ SE}} \\ \cmidrule[0.1em](l){2-5} 
 & \textbf{Cond.1} & \textbf{Cond.2} & \textbf{Cond.3} & \textbf{Cond.4} \\ \midrule[0.1em]
1 & 99.84 $\pm$ 0.08 & 99.68 $\pm$ 0.17 & 99.47 $\pm$ 0.17 & 99.79 $\pm$ 0.09 \\ 
3 & 99.25 $\pm$ 0.22 & 99.78 $\pm$ 0.15 & 99.89 $\pm$ 0.11 & 99.78 $\pm$ 0.15 \\ 
5 & 99.67 $\pm$ 0.18 & 99.44 $\pm$ 0.31 & \textbf{100.00 $\pm$ 0.00}\textnormal{\textsuperscript{*}} & 99.55 $\pm$ 0.20 \\
7 & 98.96 $\pm$ 0.38 & 99.31 $\pm$ 0.23 & 99.65 $\pm$ 0.19 & 99.65 $\pm$ 0.19 \\ \bottomrule[0.2em]
\end{tabular}%
}
\end{table}


\begin{figure}
    \centering
    \includegraphics[width=1\columnwidth]{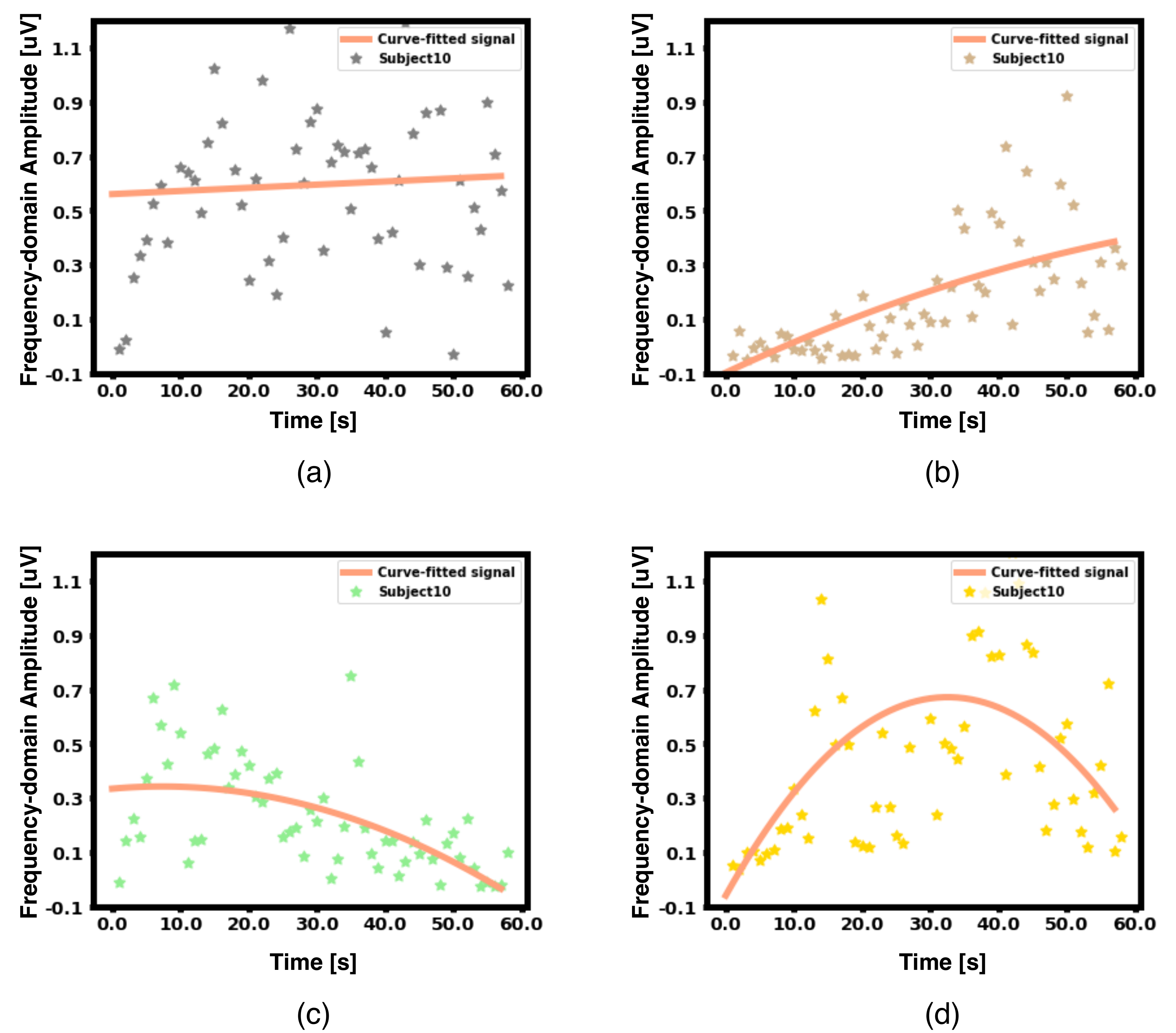}
    \caption{Curve-fitted signal of using curve fitting with polynomial functions (\textit{poly1} for cond.1 and \textit{poly2} for the rest of the conditions) on SSVEP magnitude signals. (a)--(d) display the curve-fitted of condition 1-4, respectively. Results from \textcolor{black} {the 10\textsuperscript{th} subject} of training set are presented and used as target signals to train predictive models for predicting amplitude modulations of the SSVEP signal}
    \label{Curve_fitting_target}
\end{figure}

\begin{figure*}
    \centering
    \includegraphics[width=2\columnwidth]{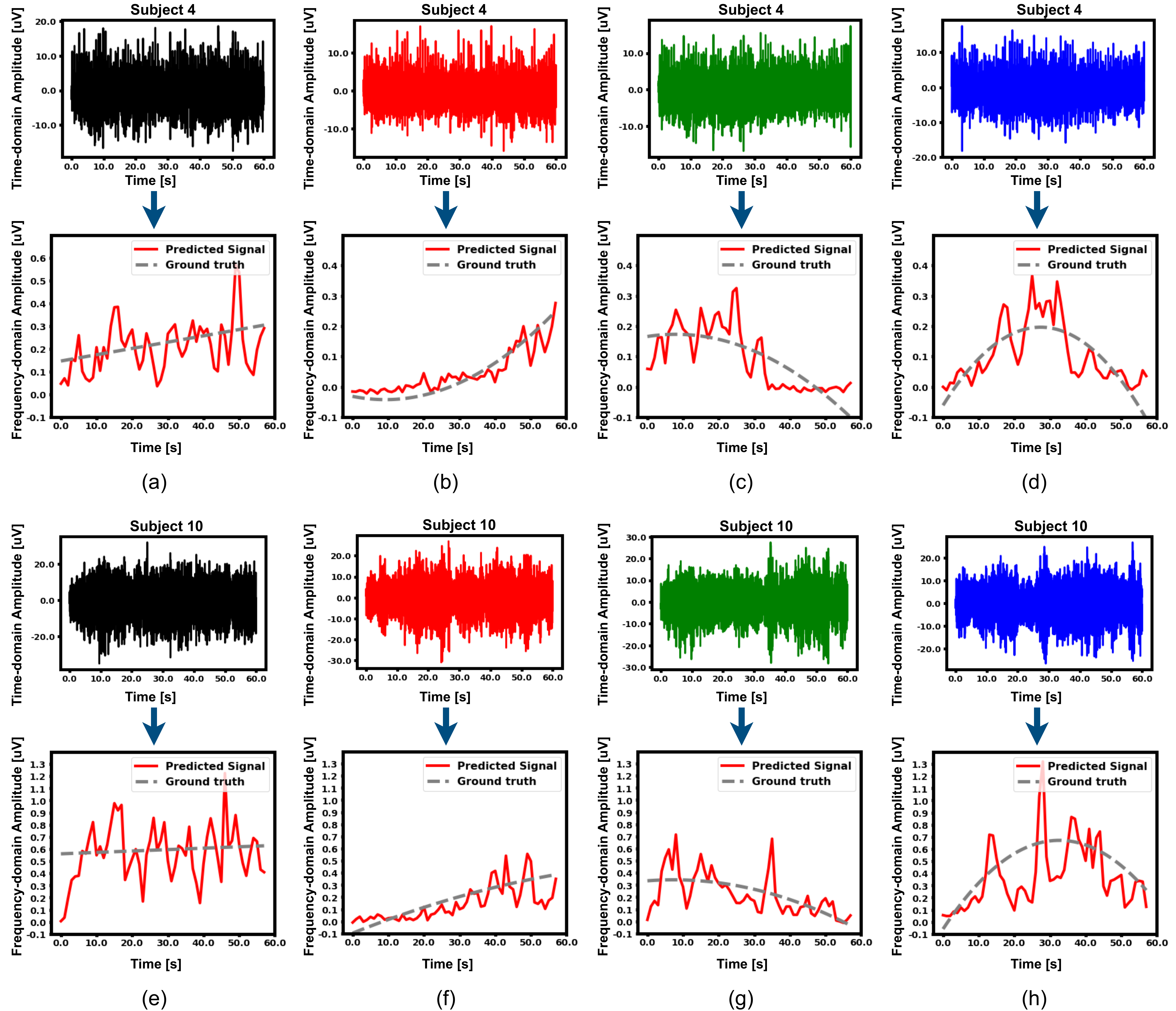}
    \caption{The 1\textsuperscript{st} and 3\textsuperscript{rd} rows demonstrate raw SSVEP responses before feeding into the proposed predictive model. The 2\textsuperscript{nd} and 4\textsuperscript{th} rows display a comparison of ground truth (target signals from performing the curve fitting method on the SSVEP amplitude signals of testing set) and predicted signals (a)-(d), (e)-(h) from experimental condition 1-4 respectively. Leave-one-subject-out cross-validation (one out of sixteen) was used to evaluate the model. \textit{Note: The results were consistent in all subjects.}}
    \label{Show_predicted_signals}
\end{figure*}
\begin{table}[]
\centering
    \caption{Comparison of average (n = 16) mean absolute error (MAE) among the four predictive models (bold is lower). * Denotes that the number is significantly lower than the others, $p < 0.05$.}
    \label{part2acc}
    \Large
    \resizebox{1\columnwidth}{!}{%
        \begin{tabular}{@{}ccccc@{}}
        \toprule[0.2em]
        \multirow{2}{*}{\textbf{\begin{tabular}[c]{@{}c@{}}Experimental \\ Conditions\end{tabular}}} & \multicolumn{4}{c}{\textbf{Avg. MAE $\pm$ SE}} \\ \cmidrule[0.1em](l){2-5} 
         & \textbf{LR} & \textbf{kNN} & \textbf{RF} & \textbf{SVR} \\ \midrule[0.1em]
        Cond.1 & 0.4583 $\pm$ 0.1464 & 0.2318 $\pm$ 0.1182 & 0.2902 $\pm$ 0.1420 & \textbf{0.1400 $\pm$ 0.0501*} \\ 
        Cond.2 & 0.5690 $\pm$ 0.2529 & 0.3035 $\pm$ 0.2074 & 0.3877 $\pm$ 0.2303 & \textbf{0.1998$ \pm$ 0.1275} \\ 
        Cond.3 & 0.4473 $\pm$ 0.1344 & 0.1876 $\pm$ 0.0894 & 0.3231 $\pm$ 0.1252 & \textbf{0.1272$ \pm$ 0.0456*} \\ 
        Cond.4 & 0.4616 $\pm$ 0.1934 & 0.2953 $\pm$ 0.1714 & 0.3492 $\pm$ 0.1939 & \textbf{0.1725$ \pm$ 0.0822} \\ \bottomrule[0.2em]
        \end{tabular}%
}
\end{table}

\begin{figure*}
    \centering
    \includegraphics[width=1.0\textwidth]{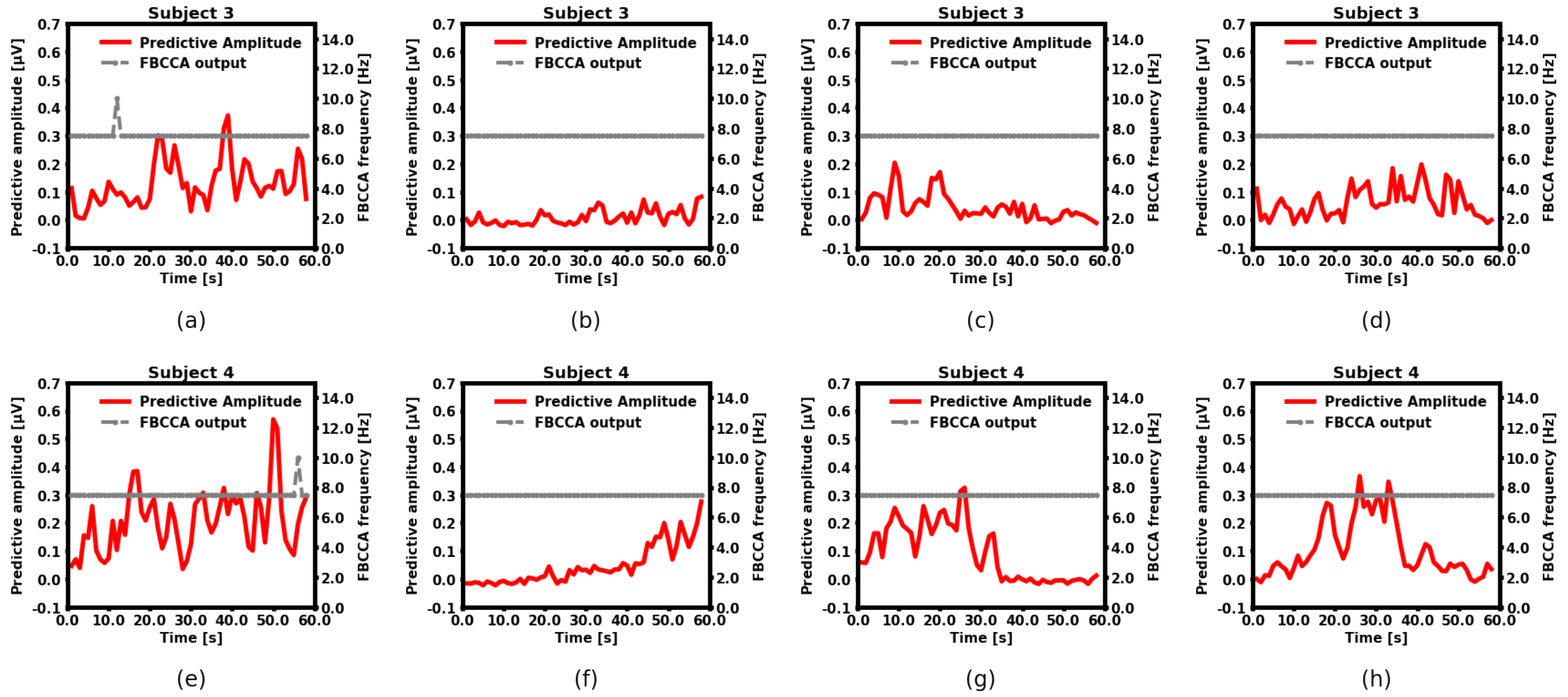}
    \caption{Results of the predicted signals ($\mu$V) and FBCCA output (Hz) of the 3\textsuperscript{rd} (upper) and 4\textsuperscript{th} (lower) subject, with the processing window length of 3 seconds and a step of 1 second. (a)--(d), (e)--(h) represent the predicted signals of condition 1--4 respectively. \textit{Note: The results were consistent in all subjects.}}
    \label{freq_predict}
\end{figure*}

\begin{figure}
    \centering
    \includegraphics[width=1\columnwidth]{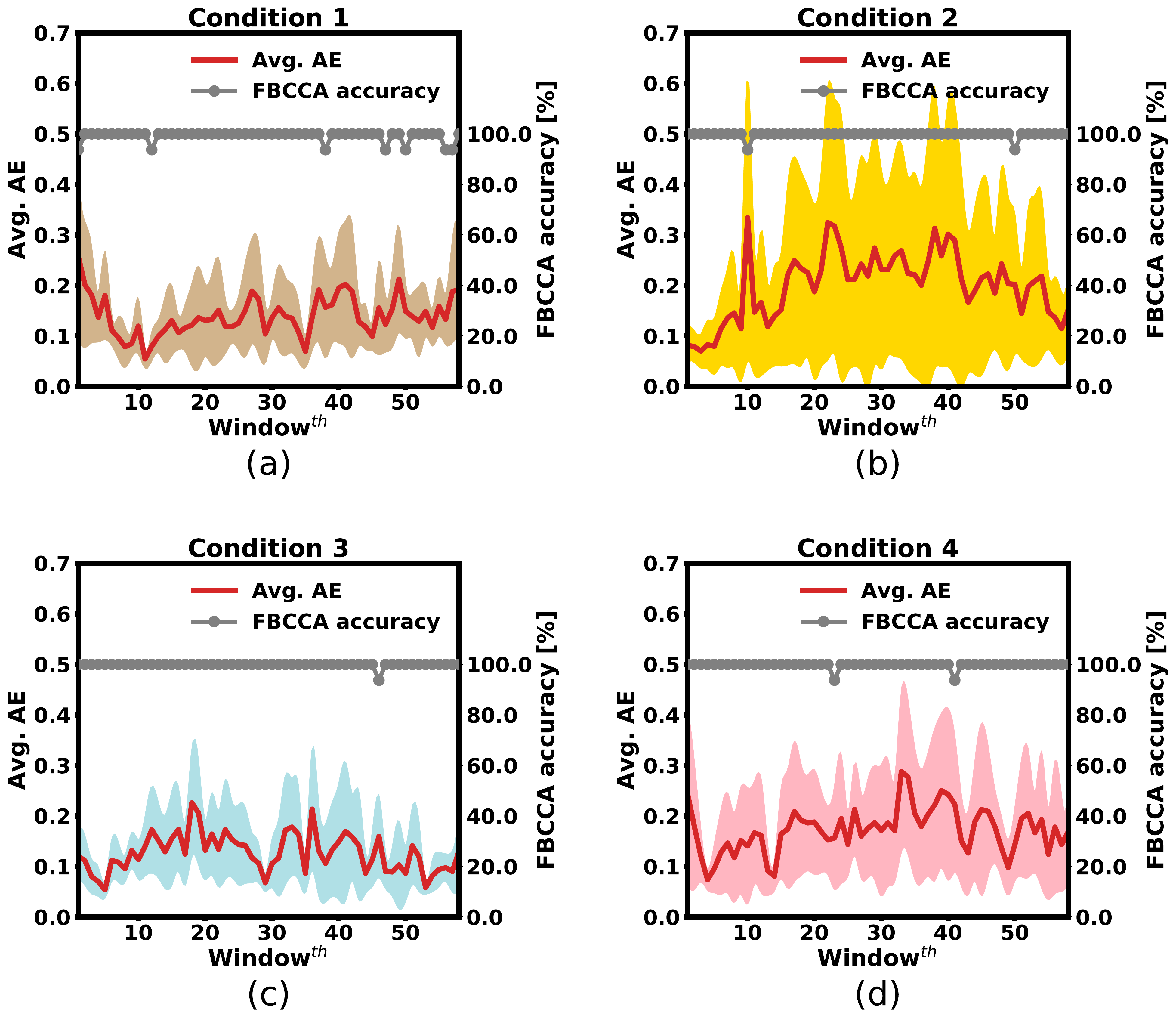}
    \caption{(a)-(d) present the averaged absolute error (AAE) and the mean of the FBCCA performance (16 subjects) from four experimental conditions. According to the results of AAE and FBCCA trade-off, the recommended contrast level for the increasing (condition 2) and decreasing (condition 3) stimulus designs are 82.35-–100.00\% and 91.76–-63.53\%, respectively (in the scale of 0–-100\% on the contrast).}
  
    \label{fbcca_mae}
\end{figure}

\subsection{Result II: Predictive Model for SSVEP Magnitude Modulation}

The purpose of this study is to identify the most appropriate predictive model to predict amplitude modulations of the SSVEP signal. \textcolor{black}{The performance of four predictive models including \textit{LR} (used for the baseline), \textit{$k$-NN}, \textit{RF} and \textit{SVR} were compared here. \autoref{part2acc} displays the average MAE of all predictive models across the experimental conditions. Each condition, the lowest average MAE was obtained once the SVR with the radial basis function (RBF) kernel was used. Descriptive statistics were calculated for all predictive models used in the study, using the ANOVA test with the Greenhouse-Geisser correction, where all the differences in performance were statistically significant in the experimental condition 1, 3 and 4. (cond.1: $F(1.464, 21.953) = 7.082$, cond.3: $F(1.314, 19.714) = 10.444$, cond.4: $F(1.112, 16.679) = 4.694$, $p < 0.05$ for all conditions). Pairwise comparisons revealed that the average MAE of the RBF-based SVR is significantly lower than the LR for both of the condition 1 and the condition 3, $p < 0.05$, but has not displayed significant differences for the $k$-NN and the RF. Overall, extensive results carried out display that the SVR is the most appropriate model for predicting the SSVEP amplitude modulations, which was included in the designed structure of this study.} 

For qualitative results, \autoref{Curve_fitting_target} displays an example of the curve-fitted signals for each experimental condition that were used as target signals for training the predictive models. In addition, the predicted signals from the SVR model were plotted per time step for each experiment condition as displayed in \autoref{Show_predicted_signals}. From this observation, we found concretely that the conventional SSVEP stimulus (cond.1) with the luminance contrast remained constant cannot be used in guiding the subjects to maintain the predictive SSVEP amplitude at a constant level. This fact is well illustrated in \autoref{Show_predicted_signals}. On the other hand, the predictive SSVEP amplitude from the proposed stimulus design (cond.2-4: the luminance contrast increased gradually, decreased gradually, or increased and then decreased) that is displayed in \autoref{Show_predicted_signals} (b)-(d) and (f)-(h), as well as the proposed stimulus design can be used as a guidance for the subject. Moreover, the qualitative results in \autoref{freq_predict} indicate a comparison between the predicted signals and the FBCCA output in each experimental condition of two subjects. Finally, \autoref{fbcca_mae} displays a comparison between the average of absolute error (AAE) of predicted signals and the performance of FBCCA in each time step for each experimental condition of all the sixteen subjects.   


\section{Discussion}

According to the experimental results, we aim to summarize the promising aspects for the possibility in using the designed SSVEP stimulus with contrast-dependent amplitude modulations of the SSVEP signal as a part of the EEG-BCI system. Firstly, the changes of luminance contrast of the visual stimulus can help subjects to manipulate the SSVEP frequency and SSVEP amplitude modulations information. Not only did we used a single-channel EEG for the EEG recording process in this experiment, but also an open-source and low-cost consumer-grade EEG device. This may lead to a more practical application of EEG-BCI system. Subsequently, a filter-bank canonical correlation analysis (FBCCA) was used as a frequency recognition method to capture the SSVEP frequency information and classify SSVEP frequency. A conventional machine learning algorithm, named support vector regression (SVR) was proposed as the predictive model to predict the contrast-dependent amplitude modulations of the SSVEP signal. Leave-one-subject-out cross-validation (LOCV) on the SVR model with the radial basis function (RBF) kernel outperforms both the $k$-NN and RF predictive models in predicting the SSVEP amplitude modulations in terms of minimizing the mean absolute error (MAE). As displayed in \autoref{Show_predicted_signals}, the SVR model provides a continuous predicted signal compared to the ground truth signal. Especially, the proposed stimulus design from experimental condition 2, 3 and 4 can be used in guiding the subjects to modulate the SSVEP amplitude. This advantage can be used as the initial step to develop online continuous SSVEP-BCIs to bridge the gap between man and machines in the future. Therefore, we can conclude that, the SVR model is promising as a predictive model to predict the SSVEP amplitude modulations, when using the designed SSVEP stimulus for further developments in EEG-BCI. 

Our study is capable of manipulating the SSVEP amplitude modulations which were obtained from EEG recordings. According to our research findings, \autoref{freq_predict} displays the result of the frequency recognition output from the FBCCA method along with the predicted signal from our proposed SVR model. While the FBCCA performs generally well on the SSVEP frequency recognition, our predictive model performs well on the SSVEP amplitude prediction.
Moreover, from the observation as displayed in \autoref{fbcca_mae}, we can recommend the luminance contrast of the visual stimulus to be used in further development of SSVEP-BCIs in the future. We chose the processing window (which the luminance contrast varies on the window number) that performs the highest accuracy in FBCCA and the lowest AAE in the SSVEP amplitude prediction for both experimental condition 2 and 3. For the experimental condition 2, we can observe that the processing window which provides the best performance for the increasing contrast stimulation is between the 43\textsuperscript{rd} and 58\textsuperscript{th} window. The luminance contrast level of this particular processing window varies between 82.35\% to 100.00\%. Moreover, in condition 3, the best performance was displayed between 5\textsuperscript{th} and 29\textsuperscript{th} window where the luminance contrast level is equivalent to 91.76\% and to 63.53\%. This observation revealed an interesting fact about luminance contrast levels and stimulus time period. If we consider the graph of experimental condition 2, we can find that high percentage of luminance contrast level performs at a higher quality behavior and it occurred at the ultimate stimulus period. Nevertheless, the behavior of experimental condition 3 is different from experimental condition 2 because it provides the best performance at high percentage of luminance contrast level during the initial stimulus period. This contradiction motivates us to explore more on two important factors. To ensure the possibility of using the SSVEP stimulus with the recommended luminance contrast, we used the recommended luminance contrast to design the visual stimulus for the experimental condition 4 in guiding the subjects. As displayed in \autoref{fbcca_mae} (d), we found that the results of the condition 4 had a high accuracy in FBCCA and a low AAE in predicting the SSVEP amplitude modulations.    

\begin{table}[]
    \centering
    \caption{\textcolor{black}{ITRs in distinguishing between the target stimulus (7.5 Hz) and the resting state (10 Hz) of the standard CCA and FBCCA methods across all experimental conditions when the processing window length of 3 seconds with the step of 1-second was selected (bold is higher). * Denotes that the number is significantly higher than the others, $p < 0.05$}}
    \label{compare_ITR}
    \footnotesize
\resizebox{0.6\columnwidth}{!}{%
\begin{tabular}{@{}ccc@{}}
\toprule[0.2em]
\multirow{2}{*}{\textbf{\begin{tabular}[c]{@{}c@{}}Experimental\\ Conditions\end{tabular}}} & \multicolumn{2}{c}{\textbf{ITR (bits min\textsuperscript{-1}) $\pm$ SE}} \\ \cmidrule[0.1em](l){2-3} 
 & \textbf{CCA} & \textbf{FBCCA} \\ \midrule[0.1em]
Cond.1 & 40.61 $\pm$ 3.69 & \textbf{54.81 $\pm$ 0.93*} \\ 
Cond.2 & 39.87 $\pm$ 5.17 & \textbf{57.09 $\pm$ 0.62*} \\ 
Cond.3 & 40.56 $\pm$ 5.22 & \textbf{57.54 $\pm$ 0.46*} \\ 
Cond.4 & 37.64 $\pm$ 4.14 & \textbf{57.09 $\pm$ 0.62*} \\ \bottomrule[0.2em]
\end{tabular}%
}
\end{table}

\textcolor{black}{To examine the computational performance of our proposed system, a general evaluation metric designed for BCI systems named information transfer rate (ITR) was calculated in this study by determining the amount of information that is conveyed by a system's output. We mainly focused on ITRs from frequency recognition for evaluating the performance. In general, the ITR (bits/min) can be calculated by \cite{obermaier2001information}:}
\textcolor{black}{
\begin{equation}
\text{ITR} = \frac{\text{log}_2C+P\text{log}_2P+(1-P)\text{log}_2(\frac{1-P}{C-1})}{t/60}
\end{equation}
}
\textcolor{black}{where \textit{C} is the number of classes for selection, \textit{P} is the classification accuracy, and \textit{t} is the average time required for a selection (in seconds).} \textcolor{black}{According to \autoref{Comparison_CCAmethod}, the classification accuracy of the standard CCA and FBCCA approaches attained from each subject was used for ITR calculation. Note that the gaze shifting time, e.g. two seconds, was discarded from the calculation due to an off-line analysis in this study. \autoref{compare_ITR} demonstrates that the ITRs of the FBCCA method were considerably higher than the CCA method. Furthermore, the paired t-test reported statistically significant differences in the ITRs of two different methods throughout all experimental conditions, $p < 0.05$. The present findings confirm the possibility of using the designed SSVEP stimulus with contrast-dependent amplitude modulations for the SSVEPs as another important component. Overall, this is an important finding in the understanding of the practical development of an online continuous-controlled SSVEP-BCI systems in the future.}

\subsection*{Study Limitations and Future Work} 
Our study has several limitations which need to be mentioned here:
\begin{itemize}
    \item Our current study is based on gaze dependent system, which has a characteristic of controlling the machine only when a user is continuously focusing on the SSVEP stimulus. Moreover, the user must be occupied in paying attention for a long time period to fulfill his/her requirement. Thus, the user might be perturbed from eye fatigue effect \cite{Fisher2005}.
    
    \item The proposed predictive model provided non-smooth predicted signals. Smooth algorithms were not applied in this study to handle these non-smooth predicted signals. Currently, we are making preparations to develop smooth algorithms.

\end{itemize}

Consequently, we can explore some directions of future researches to overcome the limitations of the current study. Generally, using the investigations from the designed SSVEP stimulus with suitable luminance contrasts over a short time period could make SSVEP-BCI systems more practical in controlling machines. Moreover, these research findings can be considered as foundations to develop smooth algorithms to support in predicting the SSVEP amplitude modulations. To do so, this system will be able to provide smooth control machines such as accelerating or decelerating the speed of a mobile robot or a robotic arm. However, to confirm this investigative study, we should validate online/real-time experiments.

\section{Conclusion}
In this study, we explored the usability of SSVEP stimulus design based on amplitude modulations of the SSVEP signal corresponded to the luminance contrast of the visual stimulus. We created a dataset for this experiment by varying luminance contrasts of the SSVEP stimulus. For practical purposes, an open-source consumer-grade EEG device, namely \textit{OpenBCI} with a single-channel EEG ($O_{z}$) was used throughout the experiment. The frequency recognition analysis exhibited that the FBCCA method significantly outperformed the standard CCA method in classifying SSVEP frequency information. Moreover, the FBCCA method performed generally well with processing window length of three-second with a step of one-second in classifying SSVEP frequency information. The support vector regression (SVR) was then proposed as the predictive model for predicting instantaneous SSVEP amplitude modulations. We obtained promising experimental results from sixteen subjects by using leave-one-subject-out cross-validation method. The integration of FBCCA and SVR was used to simultaneously estimate the SSVEP frequency and SSVEP amplitude modulations information. The ability to predict amplitude modulations of the SSVEP signal is the supreme advantage of the designed structure. Thus, this research findings can be used as a future component of EEG-BCI systems. 

\section*{Acknowledgment}
We would like to thank Fryderyk K{\"o}gl for their assistance in data collection and Thummanoon Kunanuntakij for his work in programming.

\appendix

\section{Code and supporting materials}
{
    \textcolor{black}{
    Code examples, additional figures, and other supporting materials are available on \url{https://github.com/IoBT-VISTEC/Enhancing_Detection_of_SSVEP_and_Its_Amplitude_Modulation}.}
}

\bibliographystyle{IEEEtran}
\bibliography{Reference}
\end{document}